\documentclass[letterpaper,twocolumn,10pt]{article}
\usepackage[twoside=true, head=13pt,
     paperwidth=8.5in, paperheight=11in,
     includeheadfoot=false, columnsep=2pc,
     top=1in, bottom=1in, inner=0.75in, outer=0.75in,
     marginparwidth=2pc,heightrounded]{geometry}

\usepackage{graphicx}
\usepackage[tt=false, type1=true]{libertine}
\usepackage{mathptmx}
\usepackage{amssymb}
\usepackage[varqu]{zi4}
\usepackage[T1]{fontenc}
\usepackage{enumitem}
\usepackage[font={footnotesize},labelfont={footnotesize,bf},textfont={footnotesize,it}]{caption}
\usepackage[english]{babel}
\usepackage{graphicx}
\usepackage{graphics}
\usepackage[utf8]{inputenc}
\usepackage{amsmath}
\usepackage{tabularx}
\usepackage{array, multirow} 
\usepackage{makecell}
\usepackage{enumitem}
\usepackage{array,multirow}
\usepackage[final]{listings}
\usepackage{color}
\usepackage{dblfloatfix} % fix for bottom-placement of figure
\usepackage{etoolbox}
\usepackage{float}
\usepackage{microtype}
\usepackage{subcaption}
\usepackage{blindtext}
\usepackage{filecontents}
\usepackage{multirow}
\usepackage{booktabs}
\usepackage{siunitx}
\usepackage{pifont}
\usepackage{xspace}
\usepackage{amsmath}
\usepackage{bm}
\usepackage{etoolbox}
\usepackage{pifont}
\usepackage{arydshln}
\usepackage{bm}
\usepackage{xurl}
\usepackage{tikz}
\usepackage{stmaryrd}
\usepackage{amssymb}
\usepackage{subcaption}  % in preamble
\usepackage{algorithm}
\usepackage{algpseudocode}
\usepackage{amsmath}
\usepackage{amsthm}
\newtheorem{theorem}{Theorem}
\newtheorem{definition}{Definition}
\PassOptionsToPackage{hyphens}{url}\usepackage{hyperref}

\usepackage{float}
\usepackage{etoolbox}
\AtBeginEnvironment{thebibliography}{\raggedright}
\usepackage{url}
\usepackage{hyperref}
\usepackage{breakurl}
\usepackage{microtype}
\begin{document}

% \title{Analysis of Internet Service Provider Plans at Street Address-Level Granularity}

% \title{Broadband Affordability: Understanding Broadband Plans Offered by Major ISPs across the US}

%\title{netReplica: Addressing the domain adaptation problem for network-specific learning problems}
%\title{Right, Not Just More Data: \\
%Bridging Data Gaps for ML Domain Adaptation in Networking with \sys}
% \title{When more (of the same data) is not a solution: \\ Addressing the ML domain adaptation problem in networking with \sys}
%\title{Beyond More Data: \\
%Addressing the Domain Adaptation Problem in Network ML with \sys}
% \title{Addressing the ML Domain Adaptation Problem for Networking: \\
% Realistic and Controllable Training Data Generation with \sys}

% \title{\sys: \\
% A Programmable Substrate for Realistic and Reusable Last-Mile Data Generation}

% \title{\sys: \\
% A Programmable Substrate that Unifies Fidelity, Controllability, Diversity, Composability, and Replicability in Last-Mile Data Generation}

\title{\sys: Privacy Preserving Network Verification System}  
\author{
{\rm Jaber Daneshamooz}\\
University of California, Santa Barbara\\
{\tt jaber@ucsb.edu}
\and
{\rm Melody Yu}\\
Massachusetts Institute of Technology\\
{\tt yumelody@mit.edu}
\and
{\rm Sucheer Maddury}\\
Cornell University\\
{\tt sm2939@cornell.edu}
}

% \author{Paper \#4}
\date{}

\newcommand{\smartparagraph}[1]{\noindent{\bf #1}\ }
\newcommand{\sys}{Seagull\xspace}
\newcommand{\att}{AT\&T\xspace}
\newcommand{\cv}{$cv$\xspace}
\newcommand{\metric}{Mbps/\$\xspace}
\newcommand{\diffcv}{$diff~cv_{max}_i$\xspace}
\newcommand{\maxcv}{$max~cv_{max}_i$\xspace}
\newcommand{\cmark}{\ding{51}\xspace}  
\newcommand{\eg}{e.g.\ }

\maketitle

\begin{abstract}
The Internet relies on routing protocols to direct traffic efficiently across interconnected networks, with the Border Gateway Protocol (BGP) serving as the core mechanism that manages routing between autonomous systems. However, BGP configurations are largely manual, making them susceptible to human errors that can lead to outages or security vulnerabilities. Verifying the correctness and convergence of BGP configurations is therefore essential for maintaining a stable and secure Internet. Yet, this verification process faces two key challenges: preserving the privacy of proprietary routing information and ensuring scalability across large, distributed networks. This paper introduces a privacy-preserving verification framework that leverages multiparty computation (MPC) to validate BGP configurations without exposing sensitive routing data. Our approach overcomes both privacy and scalability challenges by ensuring that no information beyond the verification outcome is revealed. Through formal analysis, we show that the proposed method achieves strong privacy guarantees and practical scalability, providing a secure and efficient foundation for verifying BGP-based routing in the Internet backbone.


\end{abstract}
\begin{sloppypar}
\section{Introduction}

Network configuration—the process of assigning settings and policies to routers and other network devices—is essential for maintaining reliable Internet connectivity and efficient resource utilization. Among these tasks, configuring routing protocols is particularly critical and complex, with the Border Gateway Protocol (BGP) serving as the de facto standard for interdomain routing across the Internet backbone.

In backbone networks, routers within different autonomous systems (ASes) communicate using BGP to exchange reachability information. Administrators manually define and apply BGP policies that determine how routes are selected and propagated. Because these configurations are crafted independently by each AS, inconsistencies and misconfigurations are common, leading to potentially widespread Internet outages that can severely affect online services and e-commerce operations.

BGP’s vulnerabilities—such as route leaks, hijacks, and general misconfigurations—remain a major operational concern. These errors can cause service disruptions, traffic detours, and even information exposure to untrusted entities~\cite{10.1145/316194.316231}. Reports such as Alibaba’s 2016–2017 analysis indicate that 66\% of network incidents stem from misconfigurations, with 56\% attributed to configuration updates and 10\% to hardware-triggered bugs~\cite{10.1145/3229584.3229585}. In 2019, a Verizon misconfiguration redirected large volumes of web traffic to unintended paths~\cite{mccarthy2019verizon}, while Meta’s 2021 outage cost an estimated \$100 million in lost advertising revenue~\cite{morris2021facebook}. The financial toll of such failures is steep, with average downtime costs reaching roughly \$9,000 per minute~\cite{ponemon2016cost}.

To prevent these incidents, network verification techniques are used to validate configuration correctness before deployment. However, verification introduces critical privacy challenges. The process requires access to sensitive configuration data—such as routing policies and topology information—that operators are reluctant to share. This concern becomes especially acute in \textbf{inter-AS verification}, where multiple independent networks must cooperate to verify global properties like loop-freedom or reachability.

Within a single AS (intra-AS verification), privacy is less of an issue since all data resides under one administrative domain. In contrast, inter-AS verification demands collaboration among independently managed ASes that treat their configurations as proprietary and confidential, often due to security and business sensitivities. Hence, a practical verification system must ensure strong privacy guarantees while enabling collective validation.

To address these challenges, this work presents \sys , a privacy-preserving framework for inter-AS network verification based on secure multiparty computation (MPC). \sys  allows multiple ASes to jointly verify network properties—such as loop-freedom—without revealing their private configurations.

Section~\ref{chap:background} introduces the background and problem formulation. Section~\ref{chap:methodolgy} presents \sys ’s design and methodology. Section~\ref{chap:threat} outlines the threat model and privacy guarantees, followed by Section~\ref{chap:eval}, which evaluates performance and scalability. Finally, Section~\ref{chap:related} discusses related work and its limitations.

\section{Background}
\label{chap:background}

\begin{table*}[t]

\centering
\caption{BGP Table Example}
\label{tab:bgp_sample}
\begin{tabular}{l l r r r l}
\toprule
\textbf{Network} & \textbf{Next Hop} & \textbf{Metric} & \textbf{Local Preference} & \textbf{Weight} & \textbf{Path} \\
\midrule
8.8.8.0/24      & 203.0.113.1  & 0 & 100 & 32768 & 15169 i \\
192.0.2.0/24    & 198.51.100.2 & 0 & 200 & 30000 & 3356 15169 i \\
198.18.0.0/15   & 198.51.100.4 & 0 & --  & 15000 & 701 1299 6453 i \\
203.0.113.0/24  & 198.51.100.3 & 70 & --  & 25000 & 174 6939 i \\
\bottomrule
\end{tabular}
\end{table*}

\begin{figure}[]
    \centering
    \includegraphics[width= 8cm]{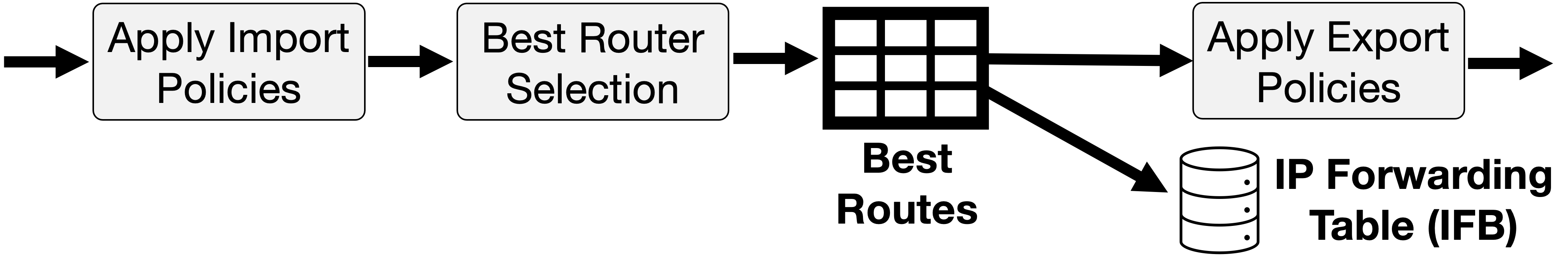}
    \caption{BGP route computation process}
    \label{fig:bgp_route}
\end{figure}

\begin{figure}[]
    \centering
    \includegraphics[width = 8.9cm]{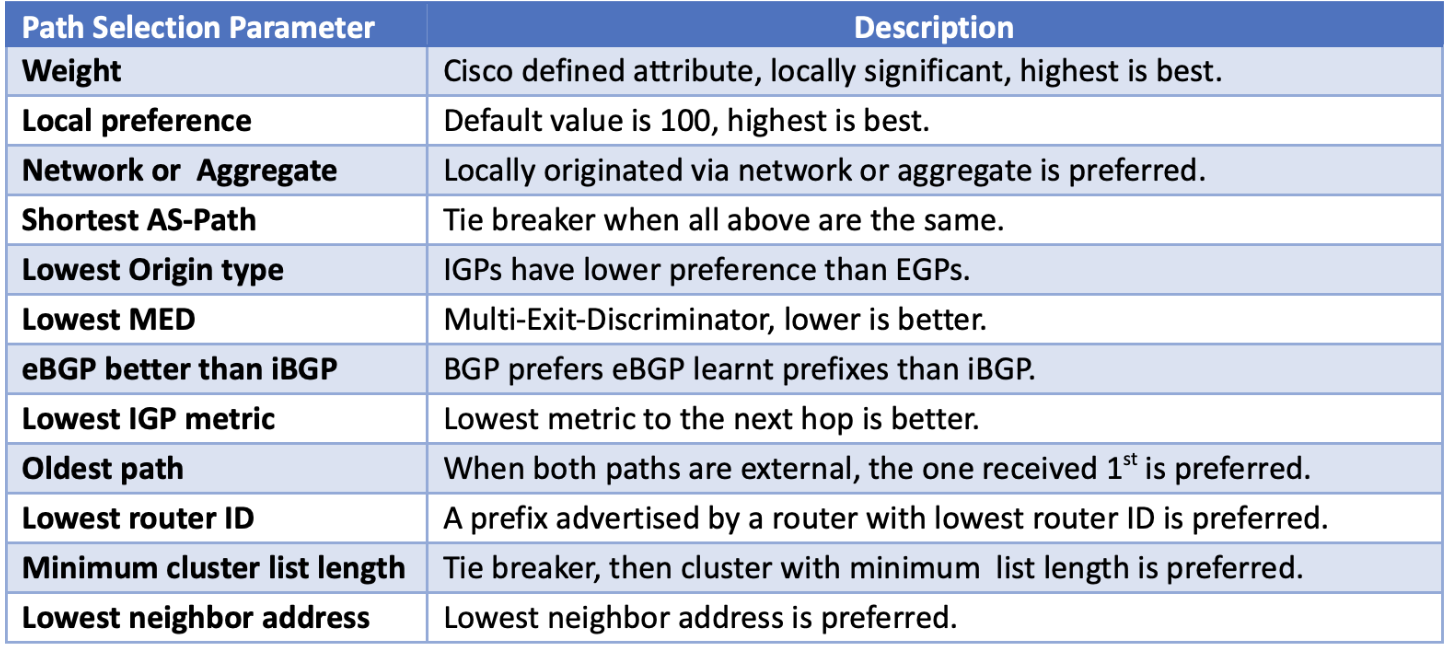}
    \caption{BGP path selection algorithm~\cite{networkslearning2020best}}
    \label{fig:policy_table}
\end{figure}

The Border Gateway Protocol (BGP) is the primary routing protocol that maintains connectivity across the Internet backbone by exchanging routing and reachability information among Autonomous Systems (ASes). Figure~\ref{fig:bgp_route} illustrates the BGP route computation process. Upon receiving BGP advertisements, the protocol applies import policies and evaluates multiple criteria to select the best routes. The Forwarding Information Base (FIB) is then populated with the selected routes, and BGP subsequently advertises updates to neighboring ASes based on export policies. BGP’s route computation and selection process depend on several factors~\cite{1208929}, including path properties (e.g., shortest {\tt AS\_PATH} and prefix length), routing policies derived from business relationships as described by the Gao--Rexford model~\cite{974523}, and control parameters such as {\tt Local\_Preference}, which enable operators to prioritize specific paths according to local policy objectives. Figure~\ref{fig:policy_table} shows the list of these path selection parameters. 

\subsection{Border Gateway Protocol}

ASes running BGP maintain three primary routing structures: the BGP neighbor
table, the BGP table (also called the Routing Information Base, or RIB), and
the BGP routing table. The neighbor table lists all connected BGP peers and
their states. The BGP table aggregates Network Layer Reachability Information
(NLRI) learned from multiple neighbors, where each destination may have several
candidate routes with different attributes (Table~\ref{tab:bgp_sample}). From this table, the router selects
the best routes based on local policies and installs them into the Forwarding
Information Base (FIB), which determines the actual packet forwarding behavior
\cite{1208929,rfc4271}.

BGP policies are central to how routes are selected and advertised, and they are strongly shaped by the commercial relationships between ASes \cite{1354576}. Two common relationships define these interactions. In a {\em customer–provider} relationship, a customer AS pays a provider AS for transit. In a {\em peering} relationship, two ASes exchange traffic between their customers at no cost, reducing transit expenses while improving performance.

\begin{figure}[ht]
    \centering
    \includegraphics[width = 8cm]{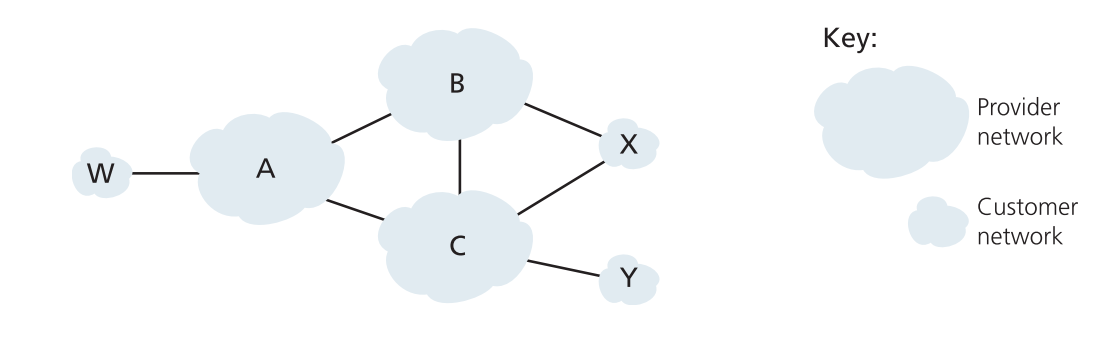}
    \caption{BGP customer-provider relation~\cite{10.5555/2584507}}
    \label{fig:customer}
\end{figure}

The Gao–Rexford rules~\cite{974523} capture the general policy framework: an AS typically prefers routes learned from customers over those learned from peers, and prefers peer routes over provider routes. Customer routes generate revenue, peer routes are cost-neutral, and provider routes incur cost. In addition, an AS usually does not export routes learned from peers or providers to other peers or providers to avoid carrying traffic without compensation. For instance, in Figure~\ref{fig:customer}, X is a customer of both B and C. Although X may learn a route to Y via C, it does not advertise this route to B. If it did, B could forward traffic to Y through X, forcing X to pay for transit~\cite{10.5555/2584507}.

These economic considerations make BGP configuration both essential and delicate. Policies are applied independently within each AS and are often adjusted manually for security, load balancing, or commercial objectives. Misconfigurations or conflicting policies across ASes can lead to routing leaks, persistent loops, or large-scale instability. Ensuring correct policy interaction and validating exported routes are therefore key to maintaining stable inter-domain routing.

\subsection{FIB}

The Forwarding Information Base (FIB) captures the final routing decisions made by an AS. It maps destination prefixes to the next-hop AS responsible for reaching those prefixes, as shown in Table~\ref{ASFib_bg}. For any destination $d$, the AS forwards packets to the next-hop AS recorded in its FIB.

Routers construct the FIB by running routing algorithms that select preferred paths according to local configuration and routing policies~\cite{10.1145/263105.263133}. The AS-level FIB thus reflects how IP prefixes are translated into next-hop forwarding actions. For instance, traffic destined for the IP address \texttt{184.130.10.20} would be forwarded to AS~43127, demonstrating how routing decisions are realized in practice.

Because FIB entries are derived from BGP route selection, changes in BGP policies directly affect the FIB. BGP policies specify route preferences (e.g., customer routes over peer or provider routes), while the FIB holds the concrete forwarding outcome (e.g., the chosen next-hop AS). Therefore, policy conflicts at the BGP level manifest as inconsistencies or unintended behavior in the forwarding tables.

\begin{table}[!htb]
      \caption{Network layer FIB at AS level}
      \label{ASFib_bg}
      \centering
        \begin{tabular}{|c|c|}
\hline
\textbf{Prefix} & \textbf{Next Hop(ASN)}  \\ 
\hline
\hline
25.8.0.9/16 & 30125\\ 
\hline
12.244.0.0/24 & 870\\ 
\hline
184.0.0.0/8 & 43127\\ 
\hline
128.69.16.0/24 & 870\\ 
\hline
        \end{tabular}
\end{table}

\subsection{Existing Problems}
BGP misconfigurations introduce several critical challenges in inter-domain routing, including loss of reachability, persistent routing loops, route leaks, and prefix hijacking. These issues threaten both the stability and security of the global routing system, making their detection and mitigation essential.

\smartparagraph{Reachability}
At the inter-AS level, reachability refers to the ability of traffic originating in one AS to successfully reach a destination prefix in another. Misconfigurations may prevent an AS from learning or propagating a usable route, resulting in partial or complete loss of connectivity. In this work, we define reachability as the ability of a source AS $s$ to forward traffic to a destination prefix $d$ under the current routing configuration.

\begin{figure}[H]
    \centering
    \includegraphics[width=8cm]{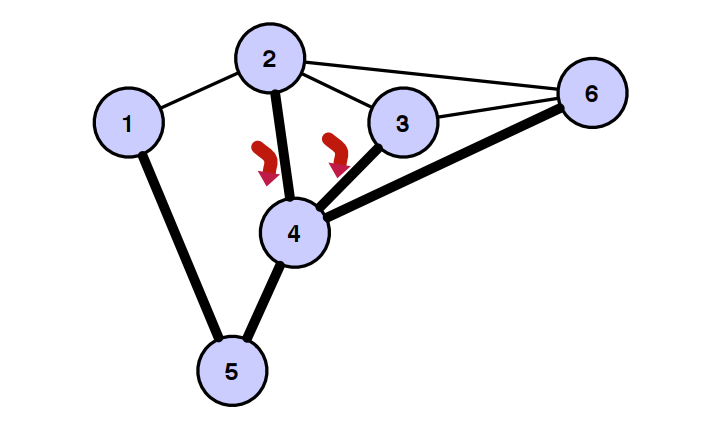}
    \caption{Policy change causing BGP loops at 2 and 3 when 4
withdraws a prefix from 2 and 3 but not 6 \cite{10.5555/1387589.1387614}}
    \label{fig:bgploop}
\end{figure}

\smartparagraph{Loop freedom}
Routing loops in inter-AS networks commonly arise from BGP policy changes. In such cases, packets circulate indefinitely among ASes until their Time-To-Live (TTL) expires, causing them to be dropped and preventing the destination from being reached. Figure~\ref{fig:bgploop} illustrates how a policy change can lead to a routing loop. AS~4 may withdraw a prefix from AS~2 and AS~3 while continuing to advertise it to AS~6, for example to shift inbound traffic from AS~2 to AS~6 in order to reduce internal congestion. However, AS~2 and AS~3 may each still believe the other has a preferred path to AS~4 based on previous advertisements. If both AS~2 and AS~3 prefer the path via each other over the path via AS~6, a routing loop forms: each forwards traffic to the other, preventing packets from ever reaching AS~4~\cite{10.5555/1387589.1387614}.

\smartparagraph{BGP route leak}
A route leak occurs when an AS unintentionally propagates routes beyond their intended export scope, violating business-relationship policies~\cite{rfc7908}. This can redirect traffic through undesired ASes, enabling observation, performance degradation, or accidental congestion. While leaks may be accidental, their impact can be equivalent to malicious interference.

\smartparagraph{Prefix hijacking}
Because BGP does not authenticate the origin of prefix announcements, an AS may falsely announce ownership of a prefix it does not control~\cite{10.1145/1282380.1282411}. This redirects traffic toward the announcing AS, enabling blackholing or interception. Recent measurements show that prefix hijacking remains frequent; for example, in early 2020, an average of 14 hijack events occurred per day~\cite{internet2020prefix}.

% \begin{figure}[hbt]
%     \centering
%     \includegraphics[width = 8cm]{Images/hijack.png}
%     \caption{Number of prefix hijacks from Jan to Jul 2020 \cite{internet2020prefix}}
%     \label{fig:hijack}
% \end{figure}

\smartparagraph{Waypoint identification: }Waypoint identification determines whether a specific AS lies on the forwarding path from a source $s$ to a destination $d$. While not itself an error condition, it is essential for tasks such as enforcing traffic inspection requirements or verifying compliance with routing policies.

\subsubsection{Contribution}

The primary contribution of this paper is the design of a network verification system that checks key routing properties in a scalable and privacy-preserving manner. The system verifies whether the network remains loop-free, whether a source AS can reach a given destination, whether a particular AS lies on a forwarding path, and whether routing announcements adhere to expected origin and export policies (i.e., no prefix hijacking or route leaks). By evaluating these properties before deployment, operators can detect configuration errors that would otherwise disrupt connectivity or violate routing agreements.

Because BGP policies reflect commercial relationships, AS operators are often reluctant to share them, even when collaboration could prevent misconfigurations. The risk of exposing business arrangements or routing strategy makes privacy a central requirement. Our system addresses this constraint by enabling verification without revealing internal policies or routing state beyond what is necessary. In doing so, the system provides a practical mechanism for assessing the safety of configuration changes while preserving confidentiality, thereby improving the reliability and security of inter-domain routing.

\section{Methodology}
\label{chap:methodolgy}

This section provides an overview of our verification system and its main components. We describe how candidate MPC frameworks were selected and evaluated, how AS-level FIBs are represented using additive secret sharing, and how verification is performed over the shared forwarding state. We then distinguish between two classes of verification tasks: \textbf{basic verification tasks} such as reachability, waypoint detection, and prefix hijacking checks, which can be evaluated directly and seamlessly from the shared FIB representation, and \textbf{loop detection}, which requires a more involved procedure. 

We began by reviewing the Awesome-MPC repository \cite{rotaru2024awesomempc}, which contains 25 MPC frameworks, and filtered for those supporting two-party computation with an accompanying paper or complete documentation. Table \ref{tab:mpc-fws} lists the shortlisted frameworks. We selected \textbf{Scale-Mamba} due to its efficient arithmetic-based secure computation suitable for operations on routing tables and its actively maintained ecosystem with clear documentation and tooling support.

\begin{figure}[t]
\centering
\begin{tikzpicture}[node distance={15mm}, thick, main/.style = {draw, circle}]
\node[main] (1) {$x_1$}; 
\node[main] (2) [above right of=1] {$x_2$};
\node[main] (3) [below right of=1] {$x_3$}; 
\node[main] (4) [above right of=3] {$x_4$};
\node[main] (5) [above right of=4] {$x_5$}; 
\node[main] (6) [below right of=4] {$x_6$};
%\node[main] (7) [right of=5] {$x_7$};
%\node[main] (8) [right of=6] {$x_8$};
\node[main,red] (7) [below right of=5] {$Dst$};
\draw[->,red] (1) -- (3);
\draw[->,red] (3) -- (6);
\draw[->,red](2) -- (5);
\draw[->,red] (5) -- (7);
 \draw[->,red](4) -- (5);
\draw[->,red](6) -- (7);
\draw[-] (4) -- (3);
\draw[-] (4) -- (2);
\draw[-] (1) -- (2);
\draw[-] (3) -- (2);
\draw[-] (1) -- (6);
\draw[-] (4) -- (6);
%\draw[-] (3) -- (4);
\draw [->, red] (7) edge[loop right]node{} (7);
\end{tikzpicture}
\caption{Network topology and its corresponding forwarding graph of destination Dst(annotated in red)}\label{fig:forwardGraph1}
\end{figure}
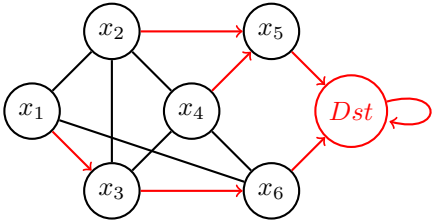

We also leverage characteristics of the internet’s topology in the design of our verification procedures. The Autonomous System (AS) graph exhibits a shallow structure, with RIPE reporting an average AS path length of 4.5 \cite{kuhne2012update}. Based on this observation, we set a traversal limit of 15 hops when evaluating reachability or detecting loops. If a path exceeds this threshold without reaching the destination, the destination is treated as unreachable. This bound ensures efficiency while remaining consistent with real-world routing behavior, and can be adjusted if needed. Together with optimizations such as pruning leaf nodes, batching operations, and reusing intermediate results, the system enables scalable and privacy-preserving verification of routing properties.

\begin{table*}[htb]
\centering
\caption{Review and comparison of MPC frameworks from Rotaru's Awesome-MPC}
\label{tab:mpc-fws}
\resizebox{\textwidth}{!}{%
\begin{tabular}{@{}lllll@{}}
\toprule
Framework      & Adversary Type                         & Platform   & Intended Use           & Method                     \\ \midrule
ABY3 \cite{10.1145/3243734.3243760}           & Semi-honest                            & -          & Secure ML              & Mixed protocol             \\
Crypten \cite{10.5555/3540261.3540640}        & -                                      & PyTorch    & Secure ML              & Secret sharing             \\
FRESCO \cite{FRESCO}         & Semi-honest/malicious                  & Java       & Data science           & Library format             \\
HoneyBadgerMPC \cite{10.1145/3319535.3354238} & Semi-honest                            & -          & Blockchain integration & Guaranteed output delivery \\
JIFF \cite{albab2019jiff}           & Semi-honest                            & JS         & Application building   & Beaver triples generation  \\
libTMCG \cite{Stamer2005EfficientEG}        & Semi-honest                            & C++        & Online card games      & Library format             \\
MOTION \cite{10.1145/3490390}         & Semi-honest                            & -          & Any                    & Mixed protocol             \\
MP-SPDZ \cite{10.1145/3372297.3417872}        & Semi-honest/malicious                  & C++/Python & Benchmarking           & Garbled circuits           \\
MPyC \cite{schoenmakers2024mpyc}           & -                                      & Python     & Any                    & Mixed protocol             \\
Rosetta \cite{Rosetta}        & Semi-honest, honest majority           & TensorFlow & Secure ML              & SecureNN, Helix            \\
SCALE-MAMBA \cite{aly2021scalemamba}    & Malicious, supports dishonest majority & -          & Any                    & Secret sharing             \\
SecretFlow-SPU \cite{288747} & -                                      & C++/Python & Secure ML              & Provability, measurability \\
Sharemind \cite{10.1007/978-3-540-88313-5_13}      & Semi-honest, honest majority           & -          & Data-driven services   & Secret sharing             \\
TF Encrypted \cite{dahl2018privatemachinelearningtensorflow}   & Semi-honest                            & Tensorflow & Secure ML              & Secret sharing             \\
TNO-MPC \cite{rooijakkers2023tno}        & -                                      & Python     & Any                    & Secret sharing             \\ \bottomrule
\end{tabular}%
}
\end{table*}

\subsection{Input Representation and Sharing}

The forwarding behavior of an AS is determined by its FIB, which reflects the outcome of its BGP routing policies. Because FIB entries encode concrete next-hop decisions, any policy conflict or misconfiguration appears directly in the forwarding tables. Seagull leverages this property by using the FIB of each AS as the basis for verification, allowing routing inconsistencies such as loops or reachability failures to be identified directly from forwarding behavior rather than relying on high-level policy descriptions.

For each destination prefix, the FIB induces a directed forwarding graph, represented as a list of edges, where each node has exactly one outgoing edge except the destination, which has none (Figure~\ref{fig:forwardGraph1}). A node with no outgoing edge indicates that the destination is unreachable from that node. If multiple next hops for the same destination are observed, only the most recent is retained.

\begin{figure}[H]
\centering
\includegraphics[width=8cm]{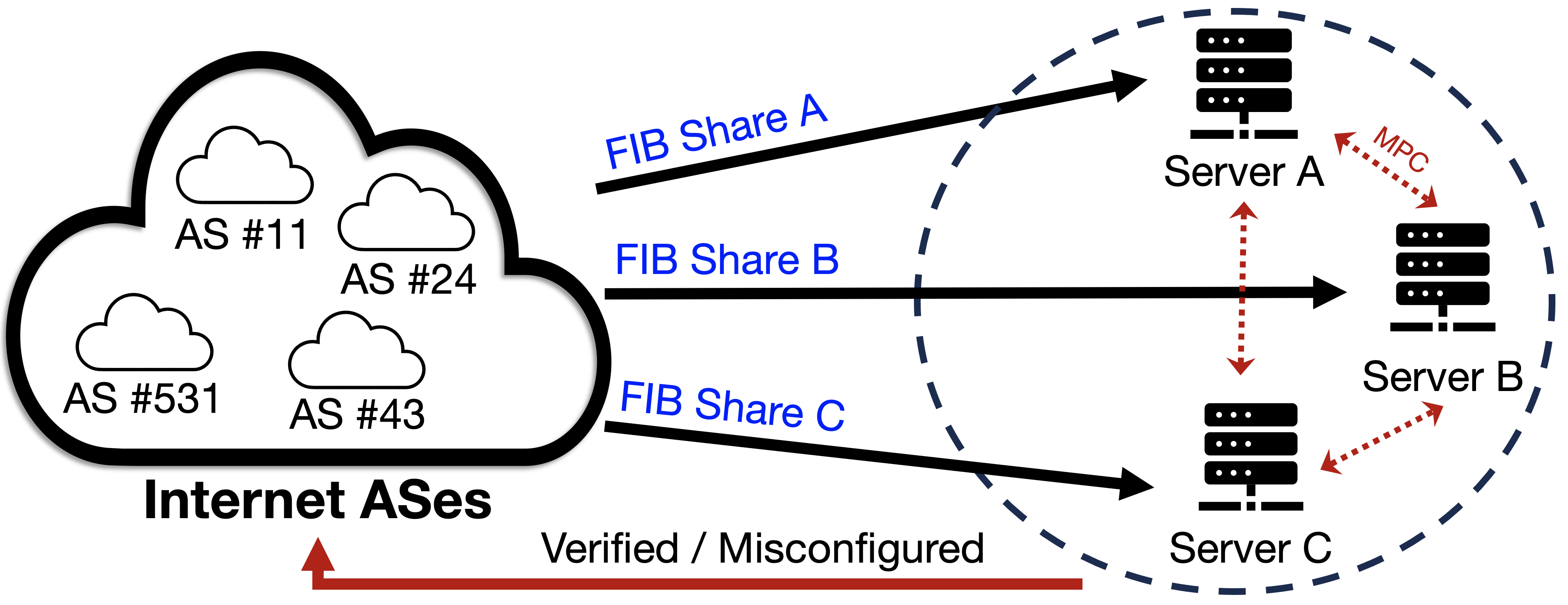}
\caption{Initial setup: ASes sharing FIB with the verifier through Additive Secret Sharing method}
\label{fig:setup}
\end{figure}

In the setup shown in Figure~\ref{fig:setup}, each AS shares its FIB with the verifier using additive secret sharing \cite{xiong2021efficientprivacypreservingcomputationbased}. During the initial setup, ASes send their full FIBs, and thereafter, only incremental updates resulting from policy changes are shared. The verifier consists of three or more servers running MPC; each FIB entry is split into shares distributed among them. No individual server learns the underlying values, and reconstruction is only possible if the servers combine their shares. This design enables Seagull to verify routing properties in a scalable and privacy-preserving manner while preventing the disclosure of routing policies or business relationships. In the following sections, we first describe the verification tasks that can be directly and seamlessly performed using the shared FIB representation. We then present the loop detection algorithm in detail.

\subsection{Basic Verification Tasks}
\label{sec:basic_verification}
\smartparagraph{Reachability}
The reachability check determines whether a destination node $d$ is reachable from a source node $s$. Using the forwarding graph representation, the verifier performs an MPC-based traversal starting at $s$ and follows the next-hop entries toward $d$. If $d$ is reached within the maximum hop bound of 15, the result is true; otherwise, the destination is deemed unreachable.

To preserve privacy, the traversal is executed in a data-oblivious manner. The verifier does not stop early when the next hop is found; instead, it scans the full array of edges at every step and always performs exactly 15 iterations before returning a result. This ensures that no access patterns reveal which entries correspond to specific nodes or routing decisions, while still allowing the reachability property to be verified securely and efficiently.

\smartparagraph{Waypoint detection} 
To verify whether a node $w$ lies on the forwarding path from source $s$ to destination $d$, we follow the same path reconstruction procedure used for reachability. As we iteratively determine the next hop from $s$ toward $d$, each node encountered is compared to $w$. If a match occurs, we conclude that $w$ is a waypoint on the path. While the protocol reveals only a yes/no result, repeated waypoint queries without restriction can leak routing structure. An adversary issuing many such queries could infer next-hop relationships for a node and eventually recover underlying routing policies. Therefore, waypoint queries should either be disabled when strict confidentiality is required, or regulated using a privacy budget that limits the number of allowable queries. This precaution prevents policy inference while still enabling waypoint verification when needed.

\smartparagraph{Prefix hijacking detection and prevention}
Prefix hijacking is a significant security concern, and Seagull addresses it using a lightweight verification approach that does not require graph traversal. In the FIB representation, the AS that owns a prefix appears as the node for which the next hop is itself. Therefore, for each prefix, we check that exactly one AS exposes such a loopback entry. If more than one AS presents itself as the origin for the same prefix, we flag a potential hijack. This verification can be performed with a single pass over the shared FIB entries.

To support this check, we also maintain a secretly shared mapping from AS numbers to their allocated IP prefixes based on Regional Internet Registry (RIR) assignments\cite{rir2024asn}. This allows the verifier to confirm whether a claimed prefix owner aligns with authoritative allocation records, without revealing either the prefixes or the AS identities in plaintext.

As a complementary protection mechanism, Resource Public Key Infrastructure (RPKI)\cite{10.1007/3-540-61770-1_45} provides cryptographic validation of prefix-to-AS bindings. Integrating or cross-referencing RPKI attestation further strengthens the detection process by ensuring that only authorized ASes can originate routes for their assigned prefixes.

\smartparagraph{BGP route leak}
BGP route leaks may arise from misconfigurations or intentional actions. Seagull can detect leaks that result in observable disruptions such as prefix hijacking, blackholing, or forwarding loops, by applying the verification tasks described earlier. In these cases, the leak manifests as a reachability or path-consistency failure, which Seagull can identify from the shared forwarding state.

However, when leaks are caused by improper export/import policies or malicious behavior that does not affect reachability (e.g., for traffic analysis), Seagull cannot reliably detect them. These cases do not alter next-hop behavior in a way that is visible to the verifier. Therefore, Seagull is most effective when the leak produces a measurable impact on forwarding behavior.

\subsection{Loop Detection}
\label{LoopSection}

\begin{figure}
\centering
\includegraphics[width=8cm]{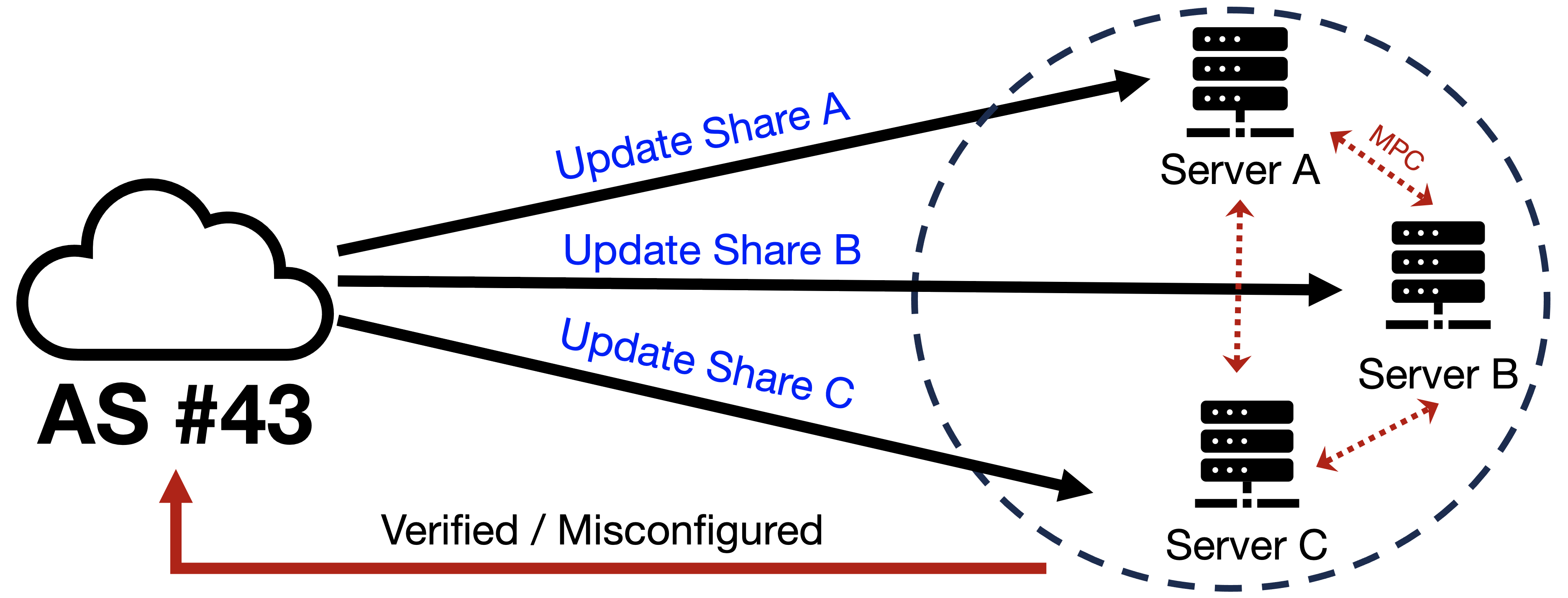}
\caption{Incremental Verification: AS~43 initiates a query to determine the safety of updating to its new policy.}
\label{fig:query}
\end{figure}

Loop detection in Seagull operates in two stages. The first stage runs when the full FIB is received for the first time. This initial verification checks global loop freedom (Figure~\ref{fig:setup}) and is the more expensive step in terms of computation and time. After this initial check, subsequent updates require only incremental verification. When a router updates its own table, it sends a query to Seagull (Figure~\ref{fig:query}). Since only one outgoing edge may change, we do not need to re-verify the entire forwarding structure. Instead, we check only the affected part of the graph, which is significantly faster. We first describe the initial verification stage, followed by the incremental query stage.

\begin{figure}[H]
\centering
\begin{tikzpicture}[node distance={15mm}, thick, main/.style = {draw, circle}]
\node[main] (1) {$x_1$}; 
\node[main] (2) [above right of=1] {$x_2$};
\node[main] (3) [below right of=1] {$x_3$}; 
\node[main] (4) [above right of=3] {$x_4$};
\node[main] (5) [above right of=4] {$x_5$}; 
\node[main] (6) [below right of=4] {$x_6$};
%\node[main] (7) [right of=5] {$x_7$};
%\node[main] (8) [right of=6] {$x_8$};
\node[main] (7) [below right of=5] {$Des$};
\draw[->,] (1) -- (3);
\draw[->] (3) -- (6);
\draw[->](2) -- (5);
\draw[->] (5) -- (7);
\draw[->](4) -- (5);
\draw[->](6) -- (1);
\draw [->] (7) edge[loop right]node{} (7);
\end{tikzpicture}
\caption{Forwarding graph with loop}\label{fig:loopg1}
\end{figure}

As previously discussed, a forwarding graph in which every node has a valid path to the destination forms a tree structure: each node (except the destination) has exactly one outgoing edge. Equivalently, the destination can be viewed as having a loopback to itself. If a set of FIB entries creates a forwarding loop, this loop appears directly in the forwarding graph. In a loop-free configuration, the forwarding graph is weakly connected; converting all directed edges to undirected edges yields a single connected component. If a loop exists, the corresponding undirected graph becomes disconnected (see Figure \ref{fig:loopg1}). Thus, verifying loop-freedom reduces to checking whether the forwarding graph is connected.

Many of the conventional loop-detection algorithms (e.g., DFS, Tarjan, Johnson) are designed for general graphs and therefore introduce unnecessary complexity for our setting. The forwarding graph already resembles a rooted tree, where each node has a single next hop toward the destination. In this structure, detecting loops reduces to verifying whether the forwarding graph is connected. To perform this check efficiently, we traverse the entire graph. Because forwarding graphs are typically wide and shallow, BFS is a natural fit. To confirm this, we benchmarked several traversal algorithms on synthetic graphs modeled after CAIDA topology data\cite{ark_ipv4_aslinks}, with $N$ nodes and $N-1$ edges. The results (Table \ref{loopbench}) validate BFS as the most effective choice for our loop-detection procedure.

\begin{table}[t]
\centering
\caption{Loop Detection Algorithms Benchmarking in Python}
\label{loopbench}
\footnotesize
\begin{tabular}{|c|c|c|c|c|c|c|} 
\hline
\textbf{\# Edges} & \textbf{BFS} & \textbf{DFS} & \textbf{Topology} & \textbf{Tarjan's} & \textbf{DSU} & \textbf{Johnson} \\
\hline
\hline
1384 & 2 & 1 & 2 & 2 & 57 & 843 \\
\hline
5500 & 6 & 5 & 7 & 7 & 46 & 732 \\
\hline
25066 & 20 & 41 & 69 & 67 & 74 & 22211 \\
\hline
25089 & 26 & 54 & 64 & 72 & 81 & 21880 \\
\hline
25071 & 23 & 49 & 59 & 69 & 111 & 22468 \\
\hline
25067 & 23 & 43 & 62 & 73 & 89 & 22757 \\
\hline
25079 & 24 & 52 & 66 & 73 & 85 & 22069 \\
\hline
\end{tabular}
\end{table}

Seagull applies BFS starting from the destination node, which acts as the root. If the forwarding graph is connected, BFS reaches all nodes. Unlike standard BFS on general graphs, we do not attempt to restart the search from another unvisited node when traversal ends. If BFS stops before all nodes are discovered, this directly indicates a forwarding loop. Since BFS proceeds opposite the direction of forwarding, we reverse the forwarding graph before traversal. This is achieved by reversing the FIB entries, as shown in Tables \ref{FIB} and \ref{RevFIB}.

\begin{table}[H]
    \begin{minipage}{.5\linewidth}
      \caption{FIB}
      \label{FIB}
      \centering
        \begin{tabular}{|c|c|}
\hline
\textbf{Source} & \textbf{Next-hop} \\ 
\hline
\hline
x1 & x2\\ 
\hline
x3 & x6 \\
\hline
x2 & x5 \\
\hline
x4 & x5 \\ 
\hline
        \end{tabular}
    \end{minipage}%
    \begin{minipage}{.5\linewidth}
      \centering
        \caption{Reversed FIB}
        \label{RevFIB}
        \begin{tabular}{|c|c|}
\hline
\textbf{Source} & \textbf{Next-hop} \\ 
\hline
\hline
x2 & x1\\ 
\hline
x6 & x3 \\
\hline
x5 & x2 \\
\hline
x5 & x4 \\ 
\hline
        \end{tabular}
    \end{minipage} 
\end{table}

The reversed FIB is privately shared among the MPC parties, with each entry representing an edge in the reversed forwarding graph and a secret-shared boolean vector $V$ indicating visited nodes. Since the destination $d$ is known, Seagull starts from $d$ and applies Algorithm \ref{alg:loop}: it scans the reversed FIB to find nodes that forward to $d$, marks them in $V$, and then iteratively repeats this process for each newly added node. The algorithm stops when either all nodes are marked (indicating a loop-free forwarding state) or no new nodes can be reached while some entries in $V$ remain false, indicating a forwarding loop. A secure aggregation over $V$ determines the final result.

\begin{algorithm}[t]
\caption{IsLoopFree($d$, $FIB$, $V$, $n$)}
\label{alg:loop}
\begin{algorithmic}[1]
\Require Destination $d$, shared boolean array $V$ of length $n$, routing table $FIB$
\Ensure Returns \textsf{True} if loop-free, otherwise \textsf{False}

\State Initialize $V[i]\gets 0$ for all $i\in\{1,\dots,n\}$
\State $V[d]\gets 1$
\State $\text{flag}\gets 1$

\While{$\text{flag}=1$}
    \State $\text{flag}\gets 0$
    \For{each \texttt{row} in \texttt{FIB}}
        \If{$V[\text{row.dst}]=1 \land V[\text{row.src}]=0$}
            \State $V[\text{row.src}]\gets 1$
            \State $\text{flag}\gets 1$
        \EndIf
    \EndFor
\EndWhile

\If{$\sum_{i=1}^{n} V[i] = n$}
    \State \Return \textsf{True}
\Else
    \State \Return \textsf{False}
\EndIf

\end{algorithmic}
\end{algorithm}

\subsubsection{Algorithm Improvements}
\label{sec:algo}
\smartparagraph{Removing leaves}
Before running the BFS, we first prune leaf nodes from the reversed FIB. In the reversed representation, each node (except the destination) has an outdegree of one. Thus, the set of nodes appearing in the second column (next-hop field) corresponds to all nodes with potential outgoing edges. If a node appears in the second column but never appears in the first column (source field), it has no children and is therefore a leaf. We remove these entries before the BFS. Since the forwarding graph is typically wide and shallow, pruning leaves substantially reduces the table size, which directly lowers the cost of privacy-preserving computation.

\smartparagraph{Discovering multiple nodes per iteration}
In standard BFS over adjacency lists, each node is dequeued and its neighbors are examined, giving a time complexity of $O(n+e)$. Here, the forwarding graph is stored as an edge list (the FIB), so repeatedly expanding neighbors one node at a time would be inefficient. Instead, in each iteration, we scan the FIB once and mark all nodes whose next hop is already marked as visited. This allows us to discover an entire frontier of nodes at once. As a result, the number of FIB scans scales with the depth of the forwarding graph (typically small, e.g., 15), rather than with the total number of nodes. This significantly improves performance in the MPC setting.

\smartparagraph{No cross edge}
In a standard BFS, the condition $x \in V$ is used to avoid revisiting nodes and forming loops. However, in our FIB, each node has out-degree 1. After reversing the FIB, each node in the resulting graph has in-degree 1. In such a graph, every edge represents a parent–child relationship, and there are no cross edges. A loop can only occur through a back edge, and in that case, the traversal will fail to reach all nodes, revealing the loop (i.e., we end up with an unreachable subgraph). Because of this structure, explicit membership checking becomes unnecessary. If the traversal reaches a point where no new node can be visited, the algorithm naturally terminates without needing an explicit $x \in V$ check. Removing this check reduces comparisons and improves efficiency while maintaining correctness.

%%%%%%%%%%%%%%%%%%%%%%%%%%%%%%%%%%%%%%%%%%%%%%%%
\subsubsection{Incremental Loop Detection (Query Phase)}
\label{sec:incremental}

Validating loop-freedom across the entire forwarding graph is expensive. As discussed earlier, identifying all nodes in the forwarding graph and re-running loop detection for every policy update is not feasible at Internet scale, where the number of ASes exceeds 100,000 \cite{rir2024asn}. Performing a full verification on each update would impose a substantial computational load on the verifier.

Additionally, the system must respond to queries within minutes. The previous loop-detection method cannot satisfy this requirement, especially because queries are processed sequentially. For example, if AS $X$ is currently being verified and AS $Y$ submits an update, $Y$ must wait until $X$'s verification completes. This queuing effect further increases latency. Therefore, we require a method that provides fast verification per update.

A forwarding graph $G$ is a tree: it is connected and contains $n$ nodes and $n-1$ edges. Suppose we add a new AS $x$ to the forwarding graph by introducing one new node and one new edge that connects $x$ to its next-hop in $G$. The resulting graph has $n+1$ nodes and $n$ edges. If the graph remains connected, then by definition, it is still a tree and therefore loop-free.

We claim that if applying a policy update by AS $x$ does not create a loop in the path from $x$ to the destination, then the entire forwarding graph remains loop-free. In the following, we justify this claim and then present a verification method that checks only the path of the AS issuing the update, enabling efficient, incremental loop-freedom verification.

%%%%%%%%%%%%%%%%%%%Union of two subgraphs%%%%%%%%%%%%%%%%%%%%%%%%%%
\begin{theorem}
\label{theorem:union}
Let $G$ and $G'$ be trees with $m$ and $n$ nodes, respectively, and thus $m-1$ and $n-1$ edges. Adding an edge $E(u,v)$ where $u \in G$ and $v \in G'$ (or vice versa) results in a connected graph with $(m+n)$ nodes and $(m+n-1)$ edges, which is also a tree. This follows directly from Kruskal's minimum spanning tree argument.
\end{theorem}

\smartparagraph{Correctness of incremental loop detection}
Consider an AS $x$ updating its routing policy. This update modifies only the next-hop of $x$, and therefore only the nodes whose paths to the destination transit through $x$ may be affected. All other nodes retain their ability to reach the destination.

We partition the forwarding graph into two subgraphs: $A$, containing $x$ and all nodes that forward to the destination through $x$; and $B$, containing the remaining nodes. Let the updated next-hop be represented as an edge $E(x,u)$, which has not yet been added to the graph.

If $u \in B$, then adding $E(x,u)$ connects the two subgraphs. By Theorem~\ref{theorem:union}, the resulting graph remains a tree and is therefore loop-free. If $u \in A$, then adding $E(x,u)$ introduces a cycle inside $A$, making the forwarding graph disconnected from the destination and creating a loop.

To determine which case applies, we either check whether $u$ is among the successors of $x$ in the reversed FIB (indicating that the update forms a loop), or verify that the new path from $x$ to the destination is reachable and loop-free using the reachability procedure described in Section~\ref{sec:basic_verification}. If the check succeeds, the update can be safely applied; otherwise, it must be prevented.

\section{Threat Model and Privacy}
\label{chap:threat}

Seagull operates under the honest-but-curious (HBC) threat model, in which parties follow the protocol correctly but may attempt to learn additional information from the data they process. A protocol is said to be $t$-private in this model if any coalition of up to $t$ parties, colluding after the protocol completes, cannot recover the private inputs. Seagull achieves $(t-1)$-privacy, where $t$ is the number of verifier servers. Thus, as long as at least one server does not collude with the others, the privacy of the data is preserved. This aligns with an anti-trust setting in which the verifier consists of mutually distrustful entities. Across the broader system, all participants are assumed to behave honestly but remain curious.

Our goal is to execute the verification algorithms in a privacy-preserving manner, ensuring that AS routing policies are not exposed. AS policies can be inferred from their Forwarding Information Base (FIB) entries; therefore, protecting the FIB during computation is essential. This requires securing the contents of the FIB as well as ensuring that access patterns do not leak information. Moreover, the final outputs of the computation must not reveal details that enable policy reconstruction. Accordingly, the only information revealed by the system is the boolean result of the verification procedure (True or False).

Seagull’s reliance on the Forwarding Information Base (FIB) for verification introduces a privacy risk. Although the FIB reveals only the best route from each AS to a destination, an active adversary can iteratively infer additional routing choices and, ultimately, the underlying BGP policies. By selectively removing best-path links and forcing BGP to recompute alternative routes, the adversary can uncover successively lower-ranked paths. Repeating this process for all neighbors of a node $x$ allows the attacker to reconstruct the relative ranking of multiple routes to a destination $D$, thereby exposing local routing policies.

Formally, let $G$ be the directed graph derived from the BGP table for destination $D$, where an edge $E(u,v)$ denotes that $v$ is the next hop from $u$ toward $D$. The FIB is a subgraph of $G$ containing only the highest-ranked edges. If the attacker gains access to the FIB, they may iteratively remove edges corresponding to selected best routes and observe which alternative edges appear next, revealing the structure of $G$ and its internal preferences. Because some BGP tables are only partially disclosed, this inference process can reveal private routing policies that ASes intend to keep confidential.

Seagull prevents this form of policy inference by ensuring that the FIB itself is never revealed during verification. The system is designed so that an adversary cannot observe routing entries, query results, or access patterns from which policy information could be deduced. Leakage risks arise at four points: data collection, storage, computation, and result sharing. Seagull’s privacy guarantees therefore require securing each of these phases so that only the final verification outcome (True/False) is revealed, without exposing any route rankings or FIB structure that could be used to infer BGP policies.

\smartparagraph{Privacy in data collection and storage}
During data collection and storage, we rely on additive secret sharing to protect confidential routing information. Although Shamir secret sharing could also be used, additive secret sharing is preferred here due to its lower computational overhead and direct suitability for arithmetic operations during later verification.

\smartparagraph{Privacy in computation}
During the computation phase, we employ multiparty computation (MPC) to ensure that only the final output is revealed. An additional requirement is that the computation must be data-oblivious.

Data-oblivious execution means that the algorithm performs the same sequence of operations and memory accesses regardless of the input. This property is essential when computation is outsourced. Although techniques such as ORAM hide the accessed index, they do not necessarily conceal the number or pattern of accesses.

\begin{definition}
Let $d$ be the input to a graph algorithm, and let $A(d)$ denote the resulting sequence of memory accesses. The algorithm is data-oblivious if, for any two inputs $d$ and $d'$ of equal length, it executes the same sequence of instructions and produces access patterns $A(d)$ and $A(d')$ that are indistinguishable to all parties.
\end{definition}

\begin{definition}
Let the forwarding graph $d$ be the input to a graph algorithm, and let $A(d)$ denote the corresponding sequence of memory accesses. The algorithm is data-oblivious if, for any two inputs $d$ and $d'$ of equal size, the instruction sequence, access patterns, and number of memory operations do not depend on the structure of the graphs.
\end{definition}

In summary, the server must not learn any information from the access pattern, including (\textit{i}) which data items are accessed, (\textit{ii}) when an item was last accessed, (\textit{iii}) whether the same item is accessed repeatedly, (\textit{iv}) whether the access pattern is sequential or random, and (\textit{v}) whether an operation is a read or a write. Our implementation, combined with Scale-Mamba's default use of ORAM, ensures that these conditions are satisfied.

\smartparagraph{Privacy in sharing the output}
The query results in our system are binary values for predefined query types. Some sensitive or repeated queries may risk leaking information. To mitigate this, we either exclude such queries or apply a privacy budget to limit repeated querying.

Because each result is a single boolean indicating whether a network property holds, it does not reveal details about the underlying FIB or policies. Since the verification system requires exact outcomes, differential privacy is not applied. This ensures both correctness and the confidentiality of the underlying data.

\section{Evaluation}
\label{chap:eval}

We implemented Algorithm~\ref{alg:loop} using the \texttt{Scale-Mamba} MPC framework and evaluated Seagull on destination-based FIBs generated from the CAIDA AS-level topology dataset (IPv4 Routed /24 AS Links Dataset~\cite{ark_ipv4_aslinks}). The experiments were conducted on a Linux machine with an Intel i7-6700K CPU (4.00GHz) and 16 GB RAM. Three MPC parties were instantiated, each holding secret shares of the FIB; all parties were executed on the same machine to avoid communication overhead unrelated to the computation itself.

Each AS in the CAIDA dataset is represented as a node labeled by its ASN, and edges indicate inter-AS connectivity. To obtain destination-based FIBs, we constructed forwarding trees using BFS. Since the AS-level topology is an unweighted connected graph, BFS yields shortest-path forwarding trees. We generated seven FIBs of varying sizes and ran the loop detection protocol privately over each.

\smartparagraph{Loop detection performance (Non-MPC)}
We implemented six Python-based loop detection algorithms and evaluated them on two types of graphs: AS-link graphs from the CAIDA dataset and destination-based FIB graphs. To examine scalability, we selected two AS-link graphs of different sizes: one with 1,384 edges and another with 5,500 edges. We also generated four destination-based FIB graphs for four randomly chosen IP prefixes. To test loop detection, we intentionally introduced loops by modifying next-hop entries. Each FIB graph contained approximately 25,000 edges.

Each algorithm was executed five times per graph, and the average execution time was recorded. Table~\ref{loopbench} summarizes the results. BFS and DFS were the most efficient traversal-based approaches. Although BFS and DFS generally exhibit similar complexity, their performance differed across graph types. On AS-link graphs, DFS performed better due to the higher variability in node degrees. In contrast, FIB graphs have at most one outgoing edge per node, which enables BFS to run more efficiently than DFS. Since Seagull operates exclusively on FIB graphs, BFS is the most suitable loop detection algorithm.

\smartparagraph{Privacy preserving loop detection}
Table \ref{SB-Loop} reports the loop detection results using SCALE-MAMBA. The final column (Query) shows the performance of incremental loop detection for a single update. The same query procedure applies to reachability and waypoint checks, since each evaluates a single path from a node to destination $d$.

In the initial phase, verifying the \textit{LoopFree} property over the entire FIB does not scale linearly. When the number of nodes doubles, the execution time typically increases by a factor of 3–4. When a loop is present, the algorithm terminates early, and the runtime depends on how soon the loop is encountered. If the loop is closer to the starting point, fewer nodes are visited, resulting in shorter execution time. For example, on a FIB with approximately 4800 nodes, full verification requires about 25 hours in the loop-free case, while detecting a loop takes about 9.5 hours. We observe the same behavior in Experiments 3–5: the total runtime depends on where the loop occurs in the graph. The earlier the loop is encountered during traversal, the fewer nodes are visited, and the faster the algorithm completes. Consequently, cases where the loop appears closer to the starting point require less computation time than cases where the loop is located deeper in the forwarding structure.

The incremental query phase remains efficient in both loop-free and loop-present cases. Since only a single path is examined, the traversal involves fewer than 15 nodes, and the runtime remains roughly constant. For instance, verifying a single policy update on a FIB with 4800 nodes completes in about 77 seconds. This demonstrates that \sys supports in-line BGP policy checks as described in Section~\ref{sec:incremental}.

These results are based on the baseline implementation of Seagull, prior to applying the algorithmic improvements in Section~\ref{sec:algo}. Even without these optimizations, the system offers practical performance in both the initial full verification phase and the incremental query phase.

\begin{table}[t]
\centering
\footnotesize % or \scriptsize
\caption{SCALE-MAMBA Loop Detection}
\label{SB-Loop}
\begin{tabular}{|c|c|c|c|c|c|}
\hline
\textbf{Exp} & \textbf{Loop} & \textbf{\#Nodes} & \textbf{\#Visited} & \textbf{Initial Phase (hr)} & \textbf{Query (s)} \\
\hline
1 & No &1245 & 1245 & 2.57 & 20.0 \\
2 & No &2708 & 2708 & 8.12 & 43.5 \\
3 & Yes &2708 & 2334 & 7.52 & 43.3 \\
4 & Yes &2708 & 2144 & 7.08 & 43.2 \\
5 & Yes &2708 & 1668 & 5.97 & 43.1 \\
6 & Yes &4800 & 1007 & 9.41 & 77.0 \\
7 & No &4800 & 4800 & 25.60 & 77.3 \\
\hline
\end{tabular}
\end{table}

\section{Related works}
\label{chap:related}
Privacy considerations have received limited attention in existing large-scale network verification systems. Even if verification is efficient, it becomes impractical when ASes are unwilling to share routing information with the verifier. Preserving privacy is therefore as critical as achieving scalability. To our knowledge, prior work does not address privacy in network verification.

Verification is computationally expensive, and applying generic privacy-preserving computation without optimization leads to prohibitive overhead. The verification algorithms must be adapted to operate efficiently under privacy constraints. Standard privacy-preserving approaches typically aim to decouple data from its owner. In network verification, this is not possible since the structure of the forwarding state inherently reflects the network’s design and policies.

Moreover, network verification requires exact correctness. The result cannot tolerate approximation. Techniques such as differential privacy, including node or edge differential privacy, introduce noise and are therefore not suitable in this setting.

Additionally, BGP includes mechanisms to avoid forwarding loops, but these mechanisms have inherent limitations \cite{abik2024network}. In eBGP, the AS-path attribute ensures that a route is discarded if an AS observes its own identifier in the path. In iBGP, the split-horizon rule prevents a route learned from one iBGP peer from being advertised to another. These mechanisms, however, do not detect or resolve the loop in the control plane; they only prevent forwarding packets along looped paths. In the eBGP case, the update is simply discarded, causing the information to be lost rather than corrected. As a result, enabling proactive detection and prevention of loops—before they affect forwarding behavior—remains necessary.

A privacy-preserving system for interdomain configuration verification is proposed in~\cite{10.1145/3600061.3600064}, enabling ASes to jointly simulate BGP configurations using data-oblivious computation. Local configurations are encoded into oblivious structures, and a converged data plane is computed via n-party SMPC. The design relies on DO-Simulation for convergence and DO-DPV for property checking. The approach incurs long verification times, supports only basic BGP features, lacks incremental verification, and cannot distinguish routing correctness from forwarding behavior.

Loop-free policy verification for SDN-enabled IXPs is explored in~\cite{10.1145/3232565.3232570}, which introduces a path verifier using Distinct-Match and SMPC to detect forwarding loops caused by BGP deflections. The scope is limited to SDXs and does not generalize to AS-level routing changes.

Logical centralization of BGP route computation using MPC is proposed in~\cite{asharov2017privacy}. The system computes routes based on business relationships or individual AS preferences using circuit-based MPC. The approach shows efficiency at regional scale but assumes a semi-honest model and lacks evidence of practicality at global scale. A related outsourcing-based approach is introduced in~\cite{10.1145/2390231.2390238}, where ASes submit routing policies to MPC servers that compute routes and return results. This design simplifies updates but faces scalability and computation overhead challenges. A reduced-information variant again in~\cite{asharov2017privacy} improves scalability by excluding stub ASes and limiting shared information to neighbor relationships and simple export preferences.

A privacy-preserving shortest-path computation using homomorphic encryption is proposed in~\cite{6733586}, where encrypted path weights are propagated to determine shortest AS paths without revealing intermediate AS identities. The main limitation is high message overhead and coordination cost.

Path validation and verification mechanisms based on cryptographic proofs are explored in several contexts. A decentralized validation protocol for multi-hop paths in 5G networks using XOR-Hash and NIZK is proposed in~\cite{9771906}, but the NIZK component introduces additional latency. A distributed fault detection mechanism for confidential systems using ZKPs is presented in~\cite{10.1145/2524224.2524233}, though practical deployment may require reducing computation and communication overheads.

Privacy-preserving route verification for wireless ad hoc networks is explored in~\cite{murugeshwari2022trust}, employing trust evaluation and private route validation. A blockchain- and TEE-based route leak protection mechanism is proposed in~\cite{9404125}, assuming limited adversarial power and sufficient honest connectivity. Hijacked route detection using IoT-based monitoring and machine learning is presented in~\cite{9474276}. Collaborative privacy-preserving node behavior verification is discussed in~\cite{pvr-ladis}. Protection against black-hole routing attacks via private path validation is introduced in~\cite{10.1007/s11276-017-1625-8}. Distributed outage diagnosis leveraging MPC is explored in~\cite{6658656,6970742}.

SAT-based interdomain configuration verification with privacy guarantees is proposed in~\cite{luo2022iveriprivacypreservinginterdomainverification}, combining Minesweeper modeling with garbled circuits, though backtracking steps reveal structural information. Static analysis for BGP configuration correctness is introduced in~\cite{10.1145/972374.972390}, but lacks privacy considerations. Symbolic route injection for verifying tagging and preference consistency is presented in~\cite{chen2021automated}, evaluated on small-scale forwarding graphs.

Using DFS and DAG are common approaches for finding loops in a graph. Running these recursive algorithms for network verification requires complete information about the graph connectivity and node policies. This issue leads to the leak of information even using privacy preserving systems. Current approaches are trying to run DFS in MPC without revealing information. Adding fake nodes and fake links to the graph is one of the approaches to hide private information in the process of running DFS in MPC. These approaches fail to keep the information private enough. Also, the process is so expensive in terms of computation and communication. Here, we present another approach to find loops in the network graph.

\section{Conclusion}
This work demonstrated that interdomain loop detection can be performed without revealing routing configurations or business relationships. By combining graph-based reasoning with secure computation, the system directly verifies whether forwarding behavior is loop-free, without relying on indirect indicators or assumptions about trust among ASes. The initial full verification phase is feasible, and the incremental query phase remains fast and scalable even in networks with 4800 ASes, making it suitable for real-time policy updates.

A current limitation is that the system focuses on a fixed routing configuration. Extending the framework to efficiently handle changes in export and import policies, as well as broader BGP decision steps, is a natural next step. Further optimization of the initial full-verification phase may also improve deployment practicality.

    \label{lastpage}

\end{sloppypar}

\small

\bibliographystyle{plain}
\bibliography{ref}

% \appendix
% \renewcommand{\thesection}{A\arabic{section}}
% \renewcommand{\thefigure}{\thesection-\arabic{figure}}
% \setcounter{figure}{0}

% \renewcommand{\thetable}{\thesection-\arabic{table}}
% \counterwithin{table}{section}

\end{document}